\documentclass[aps,twocolumn,showpacs,nofootinbib,showkeys,superscriptaddress,preprintnumbers,longbibliography,10pt]{revtex4-1}
\usepackage[utf8]{inputenc}

\usepackage{latexsym}
\usepackage{epsfig}
\usepackage{graphicx,subfigure}
\usepackage[normalem]{ulem}
\usepackage{color}
\usepackage{xcolor}
\usepackage{multirow}
\usepackage{hyperref}
\usepackage{numprint}
\usepackage{amsmath}
\usepackage{amssymb}
\usepackage{color}

\usepackage{xcolor}
\usepackage{comment}

\begin{document}

\newcommand{\SebC}[1]{{\color{red}#1}} 
\def\erm#1{\textcolor{violet}{#1}}
\def\jgb#1{\textcolor{orange}{#1}}

\def\ga{\mathrel{\raise.3ex\hbox{$>$\kern-.75em\lower1ex\hbox{$\sim$}}}}
\def\la{\mathrel{\raise.3ex\hbox{$<$\kern-.75em\lower1ex\hbox{$\sim$}}}}

\def\be{\begin{equation}}
\def\ee{\end{equation}}
\def\bea{\begin{eqnarray}}
\def\eea{\end{eqnarray}}

\def\betap{\tilde\beta}
\def\del{\delta_{\rm PBH}^{\rm local}}
\def\Msun{M_\odot}

\newcommand{\dd}{\mathrm{d}} 
\newcommand{\Mpl}{M_P} 
\newcommand{\mpl}{m_\mathrm{pl}} 

\newcommand{\CHECK}[1]{{\color{red}~\textsf{#1}}}
\newcommand{\new}[1]{{\color{blue} #1}}
\def\gmt#1{\textcolor{orange}{#1}}
\def\gmc#1{\textcolor{orange}{\tt [GM: #1]}}
 

\newcommand{\SNRmfPv}{$8.02^{+0.49}_{-0.85}$}
\newcommand{\SNRmfHPv}{$5.89^{+0.39}_{-0.91}$}
\newcommand{\SNRmfLPv}{$5.34^{+0.43}_{-0.67}$}
\newcommand{\SNRmfVPv}{$1.58^{+1.31}_{-2.24}$}
\newcommand{\PrimMassPv}{$0.62^{+0.46}_{-0.20}$}
\newcommand{\SecMassPv}{$0.27^{+0.12}_{-0.10}$}
\newcommand{\PrimSpinPv}{$0.66^{+0.13}_{-0.25}$}
\newcommand{\SecSpinPv}{$0.44^{+0.33}_{-0.39}$}
\newcommand{\ChirpPv}{$0.3527^{+0.0003}_{-0.0001}$}
\newcommand{\qPv}{$0.44^{+0.48}_{-0.28}$}
\newcommand{\MtotPv}{$0.88^{+0.35}_{-0.08}$}
\newcommand{\ChieffPv}{$0.41^{+0.08}_{-0.04}$}
\newcommand{\ChipPv}{$0.37^{+0.24}_{-0.24}$}
\newcommand{\DLPv}{$90^{+43}_{-39}$}
\newcommand{\zPv}{$0.02^{+0.01}_{-0.01}$}
\newcommand{\raPv}{$5.05^{+0.22}_{-2.08}$}
\newcommand{\decPv}{$1.13^{+0.35}_{-1.64}$}
\newcommand{\MfPv}{$0.^{+}_{-}$}
\newcommand{\MaxSNRmfPv}{$8.51$}
\newcommand{\ProbSSMonePv}{$92\%$}
\newcommand{\ProbSSMtwoPv}{$100\%$}

\newcommand{\lnBHL}{$6.27 \pm 0.10$}
\newcommand{\lnBH}{$-0.04 \pm 0.04$}
\newcommand{\lnBL}{$0.12  \pm 0.03$}
\newcommand{\lnBV}{$0.02 \pm 0.01$}
\newcommand{\lnBcohinc}{$6.17 \pm 0.11$}

\title{{Analysis of the subsolar-mass black hole candidate SSM200308} \\ from the second part of the third observing run of Advanced LIGO-Virgo} 

\author{Marine Prunier}
\affiliation{Service de Physique Th\'eorique, Universit\'e Libre de Bruxelles (ULB), Boulevard du Triomphe, CP225, B-1050 Brussels, Belgium}
\affiliation{D\'epartement de Physique, Universit\'e de  Montr\'eal (UdeM), Succ. Centre-Ville, Montr\'eal, Qu\'ebec, H3C 3J7, Canada}

\author{Gonzalo Morr\'as}
\affiliation{Instituto de F\'isica Te\'orica UAM/CSIC, Universidad Aut\'onoma de Madrid, Cantoblanco 28049 Madrid, Spain}

\author{Jos\'e Francisco Nu\~no Siles}
\affiliation{Instituto de F\'isica Te\'orica UAM/CSIC, Universidad Aut\'onoma de Madrid, Cantoblanco 28049 Madrid, Spain}

\author{Sebastien Clesse}
\affiliation{Service de Physique Th\'eorique, Universit\'e Libre de Bruxelles (ULB), Boulevard du Triomphe, CP225, B-1050 Brussels, Belgium}

\author{{Juan Garc\'ia-Bellido}}
\affiliation{Instituto de F\'isica Te\'orica UAM/CSIC, Universidad Aut\'onoma de Madrid, Cantoblanco 28049 Madrid, Spain}

\author{{Ester Ruiz Morales}}
\affiliation{Departamento de F\'isica, ETSIDI, Universidad Polit\'ecnica de Madrid, 28012 Madrid, Spain}
\affiliation{Instituto de F\'isica Te\'orica UAM/CSIC, Universidad Aut\'onoma de Madrid, Cantoblanco 28049 Madrid, Spain}

\date{\today}

\begin{abstract}

We present a follow-up study of a subsolar black hole candidate identified in the second part of the third observing run of the LIGO-Virgo-KAGRA collaboration. The candidate was identified by the GstLAL search pipeline in the Hanford and Livingston LIGO detectors with a network signal-to-noise ratio of $8.90$ and a false-alarm-rate of 1 per 5 years. It is the most significant of the three candidates found below the O3b subsolar mass false-alarm rate threshold of 2 per year, but still not significant enough above the background to claim a clear gravitational wave origin. A Bayesian parameter estimation of this candidate, denoted SSM200308, reveals that if the signal originates from a compact binary coalescence, the component masses are $m_1= \text{\PrimMassPv}M_{\odot}$ and $m_2 = \text{\SecMassPv}M_{\odot}$ (90\% credible intervals) with at least one component being firmly subsolar, below the minimum mass of a neutron star.  This discards the hypothesis that the signal comes from a standard binary neutron star. The signal coherence test between the two LIGO detectors {is consistent with, but does not necessarily imply,} a compact object coalescence origin.
\end{abstract}

\maketitle

\section{Introduction}
Since the very first detection of a gravitational wave event by LIGO in September 2015 \cite{Abbott:2016blz}, the LIGO-Virgo-KAGRA (LVK) collaboration has reported nearly a hundred gravitational-wave (GW) events from the coalescence of compact binary systems \cite{TheLIGOScientific:2016pea,LIGOScientific:2018mvr,LIGOScientific:2020ibl,LIGOScientific:2021usb,LIGOScientific:2021djp}. 
These broad-band GW detectors are able to detect a wide range of compact binary coalescences (CBC) masses and are even sensitive to the merging of hypothetical subsolar mass (SSM) compact objects $m< 1 M_\odot$. As stellar evolution models predict that neither black holes (BH) nor neutron stars can be significantly lighter than one solar mass, the detection of SSM compact objects would clearly indicate a new formation mechanism alternative to the classical scenario. The discovery of an SSM merger would therefore have revolutionary implications for astrophysics, cosmology and fundamental physics.

Several GW searches for CBCs having at least a component mass of less than 1 $M_\odot$ have been carried out using the Advanced LIGO-Virgo data~\cite{LIGOScientific:2018glc,LIGOScientific:2019kan,2021ApJ...915...54N,2021PhRvL.127o1101N,LIGOScientific:2021job,LIGOScientific:2022hai,2022PhRvD.106b3024N} with no firm detection. However, in the latest LIGO-Virgo observing run, O3b \cite{LIGOScientific:2021djp}, three candidates of SSM binary black hole events were reported \cite{LIGOScientific:2022hai}. One candidate found in O2 data~\cite{Phukon:2021cus} was also analysed in~\cite{morras_analysis_2023}. Those triggers are not classified as \textit{confirmed SSM GW events} but rather as \textit{candidate events} due to their false alarm rate (FAR) being too large to confidently claim for the existence of such revolutionary objects. However, as the sensitivity of the detectors improves and observation time is accumulated \cite{LVK_prospect_2020}, the perspectives for the future detection of an SSM compact object are hopeful.\newline

In this work, we further investigate one of these SSM triggers, the candidate event observed on March 8th 2020 $-$referred here as SSM200308$-$ reported in Table~\ref{table:search}. With a FAR of 1 per 5 years, SSM200308 is the most significant candidate of the search, found by \texttt{GstLAL} \cite{Cannon:2020qnf} in coincidence in both LIGO Hanford and LIGO Livingston detectors. Even though SSM200308 did not generate a trigger in Virgo with a signal-to-noise ratio (SNR) above the single detector threshold, Virgo was taking data at that time, which we will include in the parameter estimation (PE). We perform a follow-up of this candidate and analyze the data in detail, performing a PE of the signal. As a by-product, the PE allows us to infer the probability that the source of SSM200308 has SSM components, if one assumes that the signal comes from a binary black hole merger event.\\

The goal of this work is not to claim the detection of SSM black holes by the LVK collaboration. The possibility that the candidate is not of astrophysical origin but induced by environmental or instrumental noise cannot be excluded. 
Given the expected increase in sensitivity of future observing runs, it is important to be prepared to analyze challenging SSM signals for O4, O5, and subsequent SSM BH searches. While Ref.~\cite{Wolfe:2023yuu} explored how to do PE on zero-noise SSM injections using
Relative Binning methods~\cite{Cornish:2021lje, Zackay:2018qdy}, the present paper aims to show that a proper PE on such long duration and low mass signals can be performed in real GW data, using Reduced Order Quadrature (ROQ) methods~\cite{Smith:2016qas}. \newline

This paper is organized as follows. In Section 2, we describe the method used to perform the PE. In Section 3, we present the inferred properties of the source. In Section 4, we present the tests carried out to assess the significance of the candidate and investigate the potential nature of the source of SSM200308 before concluding in Section 5.

\begin{table}[t!]
    \bgroup
    \def\arraystretch{1.5}
	\begin{tabular}{| c | c | c | c | c |}
        \hline
        Candidate & $m_1 [M_{\odot}]$ & $m_2 [M_{\odot}]$ & SNR$_{tot}$ & FAR [yr$^{-1}$ ] \\ 
        \hline
        SSM200308 & 0.78 & 0.23 & 8.90 & 0.20 \\
        \textcolor{gray}{SSM170401} & \textcolor{gray}{4.89}  & \textcolor{gray}{0.77} & \textcolor{gray}{8.67} & \textcolor{gray}{0.41} \\
        \hline
    \end{tabular}
    \egroup
    \caption{Trigger parameters for SSM200308 as reported by \texttt{GstLAL} in Ref.~\cite{LIGOScientific:2022hai}. The trigger occurred at 2020-03-08 18:05:53. In grey we show \textcolor{gray}{SSM170401}, a candidate from O2 \cite{Phukon:2021cus} that was analyzed in detail in \cite{morras_analysis_2023}.}
\label{table:search}
\end{table}

\section{Method}
\label{sec:method}

The candidate was found in a dedicated search for GWs from compact binaries with at least one component below one solar mass performed on the Advanced LIGO-Virgo's O3b run \cite{LIGOScientific:2022hai}. The \texttt{GstLAL} pipeline reports detector frame masses of $0.78 M_\odot$ and  $0.23 M_\odot$, with a FAR of 0.20 yr$^{-1}$ and a combined network SNR of $8.90$. Given the time of O3b coincident data suitable for observation $T_{\rm obs} = 125.5$ days, the search would produce a higher-ranked candidate in $1-\exp(-T_\mathrm{obs}\cdot\mathrm{FAR})=6.5\%$ of searches on data containing only noise, assuming a Poisson distribution for the background. {This estimate ignores the possible trials factor coming from the fact that 3 search pipelines (\texttt{GstLAL}, \texttt{MBTA} and \texttt{PyCBC}) run on the data~\cite{LIGOScientific:2022hai}. Assuming these pipelines were uncorrelated and that the event was only observed by \texttt{GstLAL}, the trials factor could be of at most 3. Taking into account the trials factor, the false alarm probability (FAP) would be computed as $\mathrm{FAP} = 1-\exp(-N_\mathrm{trials} \cdot T_\mathrm{obs}\cdot\mathrm{FAR})$, which would take a value of at most $\mathrm{FAP} = 18.6\%$ in the case $N_\mathrm{trials}=3$.} \newline

In the following, we analyze SSM200308 assuming that it comes from the coalescence of two compact objects. The properties of the source are inferred by performing a Bayesian PE on the data from LIGO Livingston, LIGO Hanford, and Virgo. The strains are directly obtained from the O3b open-access data~\cite{KAGRA:2023pio}\cite{Abbott:2019ebz}. Looking at the data quality, we found only two very minor glitches, one in Livingston and one in Virgo at  226.3s and 252.0s before coalescence respectively. Even though these were not expected to significantly bias the PE, we removed them using \texttt{BayesWave}\cite{Cornish_2015,PhysRevD.91.084034,Cornish:2020dwh}. The median power spectral density (PSD) for each detector was computed from a posterior distribution of PSDs as estimated by \texttt{BayesWave}.
We choose the waveform model $\texttt{IMRPhenomPv2}$ \cite{PhysRevLett.113.151101} with spin parameters measured at a reference frequency of $f_\mathrm{ref}$ = 100 Hz to fit our candidate GW signal.\\
The priors, shown in Appendix~\ref{sec:priors}, are purposely chosen uninformative and broad on the 15 parameters, to minimize bias in PE. More precisely, we take uniform priors in component masses and spins, comoving volume, sky location and time of coalescence. The chirp mass $M_c$ was intentionally constrained in a narrow range of $M_c \in [0.351,0.355] M_\odot$ around the expected $M_c$ from the search. Indeed, we expect the chirp mass to be the best-constrained parameter as it is the dominant quantity dictating the frequency and phase evolution of the GW signal \cite{Cutler:1992tc}. Given the very small relative uncertainty that can be expected for $M_c$, it would be otherwise difficult for the nested sampler to find the extremely narrow peak in the chirp mass posterior. The final posterior distribution of the chirp mass is narrower than this prior zoom, which means that this choice does not bias the shape of the posterior distribution.\newline

The PE is performed with a template starting at 37 Hz. Assuming the chirp mass provided by the search, the signal is expected to last $\sim 245\, \mathrm{s}$. Therefore, we analyze $256\, \mathrm{s}$ of data with the help of ROQ methods \cite{Smith:2016qas}, which greatly speeds up the Likelihood evaluation time. We generate the ROQ basis in a parameter space compatible with the priors using the algorithm introduced in Ref.~\cite{Morras:2023pug}. To make sure that this method does not bias our PE, we also perform the analysis without the ROQ method on a smaller analysis duration of $120\, \mathrm{s}$ (with a template starting at 50 Hz). The two methods give similar results for the parameters of interest, i.e. the source masses, with a larger uncertainty for the first $120\, \mathrm{s}$ PE, as expected, given that we loose at most 11\% of the SNR and 14\% of the useful number of cycles $N_\mathrm{useful}$ \cite{Damour:2000gg} by starting the analysis at 50 Hz instead of 37 Hz with ROQ. A lower value of the low-frequency cut-off could have been considered, however, the SNR expected to be gained is negligible. For example, if we had used a low frequency cut-off of 30Hz, we could have gained at most 2.7\% SNR and 5.1\% of $N_\mathrm{useful}$, but the signal would be $\sim$430s, which presents some challenges to build the PSD and guarantee the quality of the data. To sample the posterior distribution we use the \texttt{BILBY} \cite{Ashton:2018jfp} Nested Sampling routine \texttt{dynesty} \footnote{The configuration files for the PE and PSDs along with PE results -corner plots and posteriors- are all available on \href{https://github.com/MarinePrunier/Analysis-Of-Subsolar-Mass-Black-Hole-Candidates-In-Advanced-LIGO-Virgo-Data.git}{github}}.

\vspace{-0.35cm}
\section{Properties of the source of SSM200308}\label{sec:properties_of_the_source}

\begin{figure}[t!]
\begin{center}
\hspace{-1cm}
\includegraphics[width=0.48\textwidth]{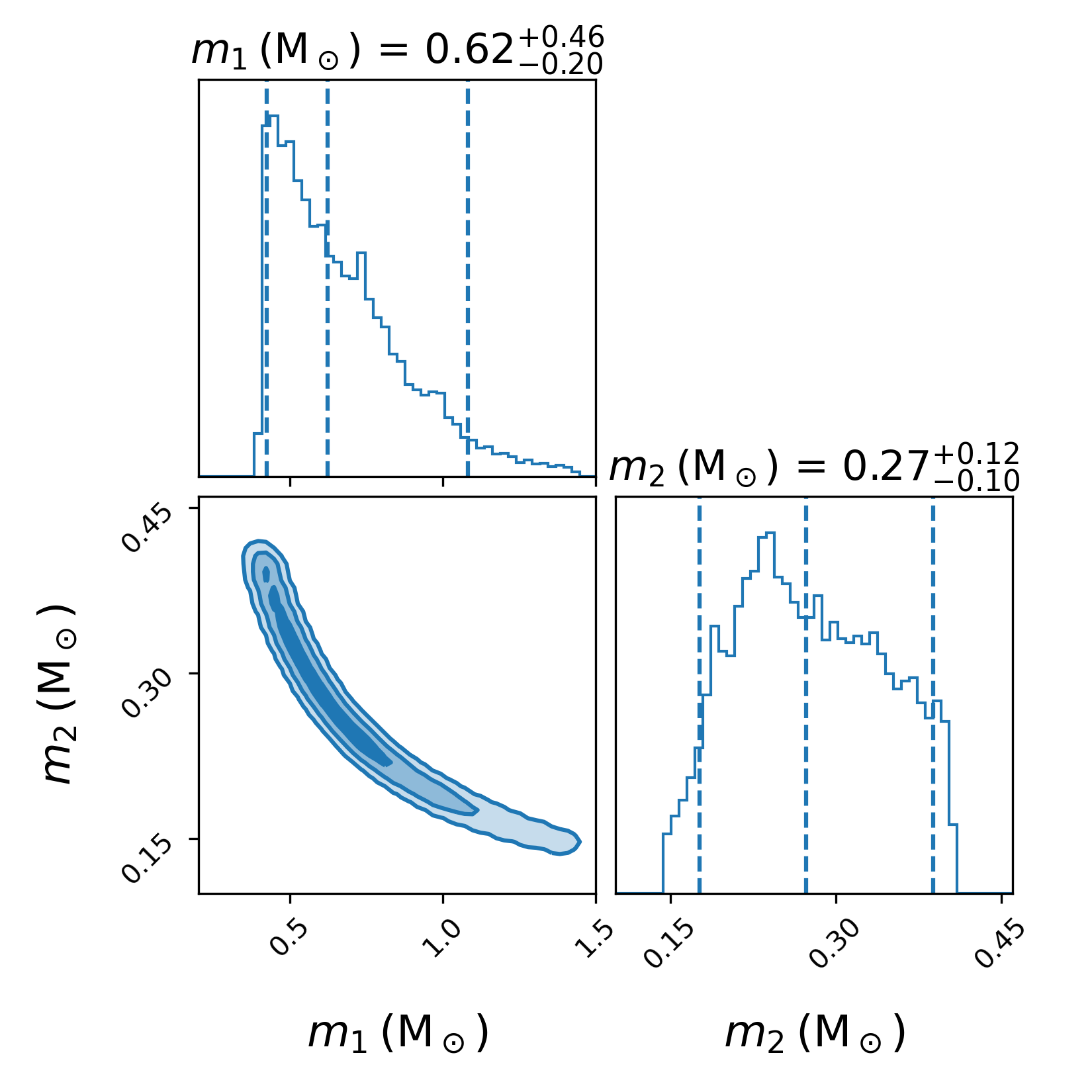}
\end{center}
\caption{Posterior distributions for the primary and secondary mass in the source frame. The 90$\%$ credible region is indicated by the outer dashed vertical lines in the marginalized distributions. The joint distribution contours denote the 39.3$\%$, 86.5$\%$ 98.9$\%$ credible regions \cite{Lista:2016chp}.}
\label{fig:lalinference_m1m2corner}
\end{figure}

\begin{figure*}[t!]
\begin{center}
\includegraphics[width=0.48\textwidth]{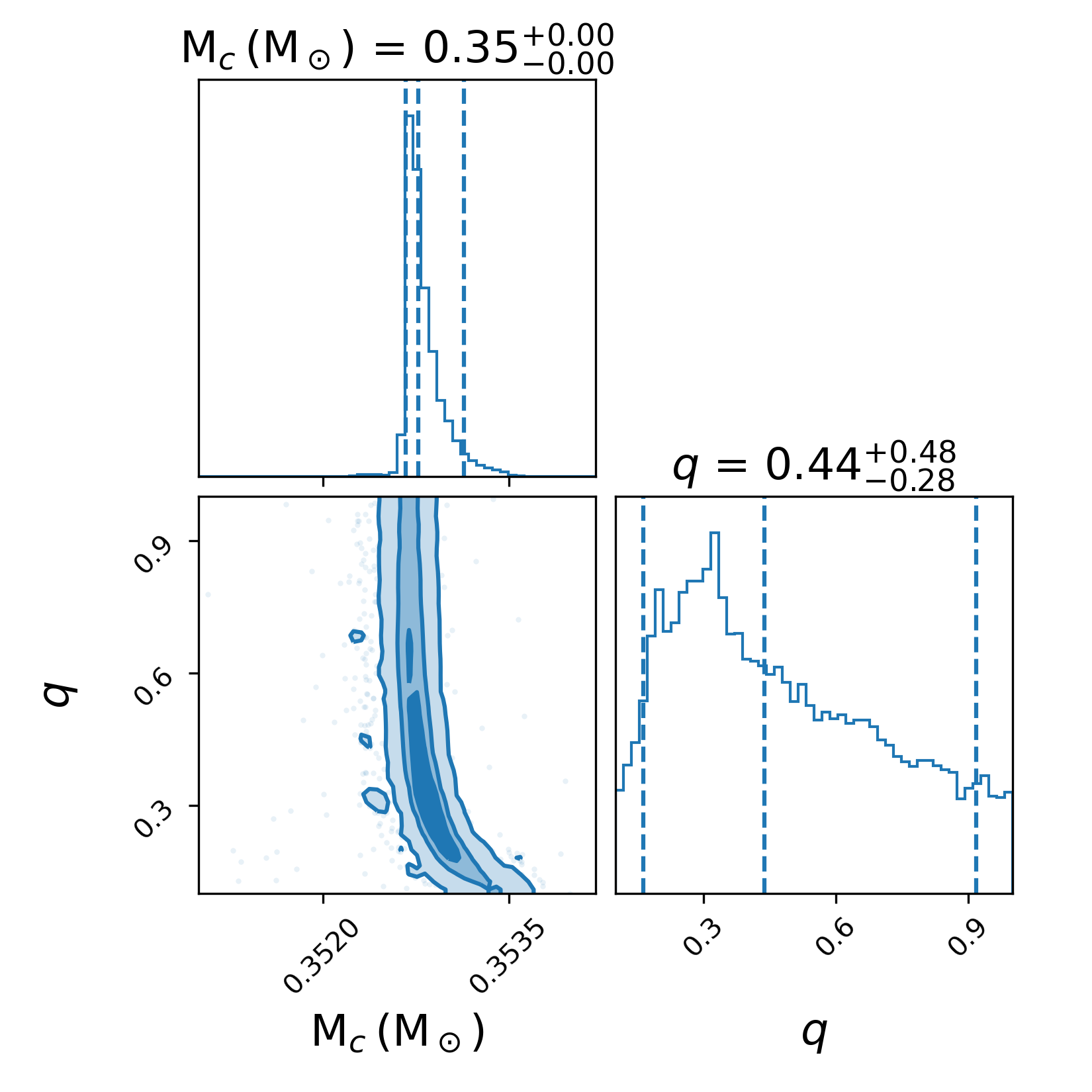}
\includegraphics[width=0.48\textwidth]{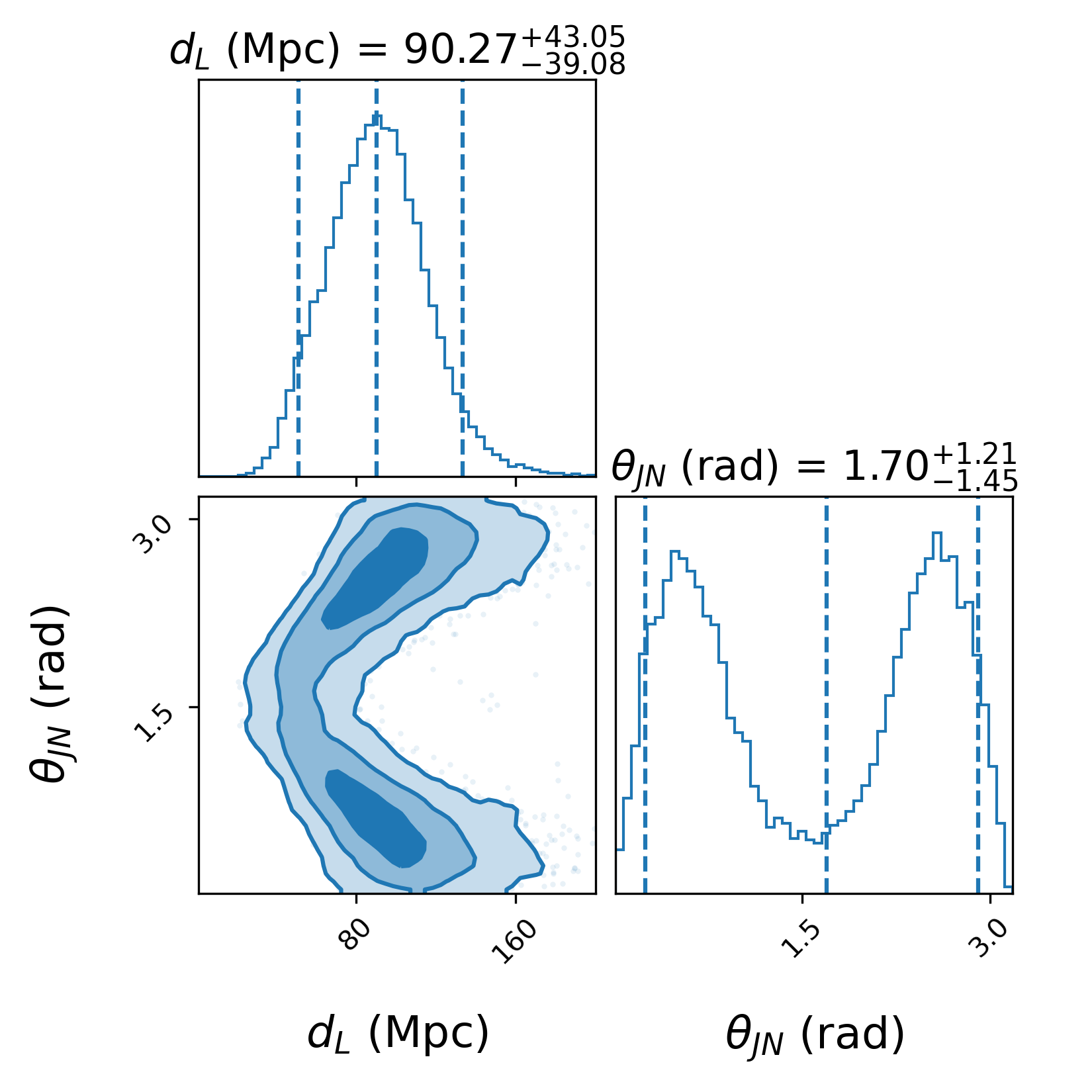}
\end{center} 
\caption{\textbf{Left:} posterior distribution for the chirp mass and mass ratio of the source of SSM200308. \textbf{Right:} posterior distributions for the luminosity distance [Mpc] and the inclination angle [rad] of the source of SSM200308. The inclination angle indicates the angle between the line of sight and the total angular momentum of the binary. For nonprecessing binaries, this is equal to the angle between the orbital angular momentum and the line of sight. The 90$\%$ credible region is indicated by the outer dashed vertical lines in the marginalized distributions. The joint distribution contours denote the 39.3$\%$, 86.5$\%$ 98.9$\%$ credible regions \cite{Lista:2016chp}.}
\label{fig:lalinference_Dist}
\end{figure*}

Table \ref{table:parameters} summarizes the values found for several significant parameters of the source of SSM200308, assuming a compact binary merger origin. It has individual source-frame masses $m_1= \text{\PrimMassPv}M_{\odot}$ and $m_2 = \text{\SecMassPv}M_{\odot}$ as shown in Fig.~\ref{fig:lalinference_m1m2corner}, for each parameter, we report the median value and the 90 \% credible interval. The marginalized posterior distribution for the first mass favors a mass lower than 1 M$_\odot$ at 92\% and for the second mass, the whole posterior distribution lies below 1 M$_\odot$. These component masses were computed using the chirp mass $M_c$ and the mass ratio $q$ (Fig.\ref{fig:lalinference_Dist}). The detector frame chirp mass is tightly constrained to be \ChirpPv~$M_\odot$, allowing us to constrain with relatively high accuracy the source properties of SSM200308 in spite of its low SNR. \newline

It is interesting to note in Table~\ref{table:parameters} that the effective inspiral spin parameter $\chi_\mathrm{eff}$ is relatively well measured and, with a value of $\chi_\mathrm{eff} = $\ChieffPv, it is found to be significantly larger than 0,  meaning that at least one component has a non-zero spin.
The posterior on the precession spin parameter $\chi_p$ is similar to the prior and therefore uninformative. 

\begin{figure}[ht]
\begin{center}
\includegraphics[width=0.5\textwidth]{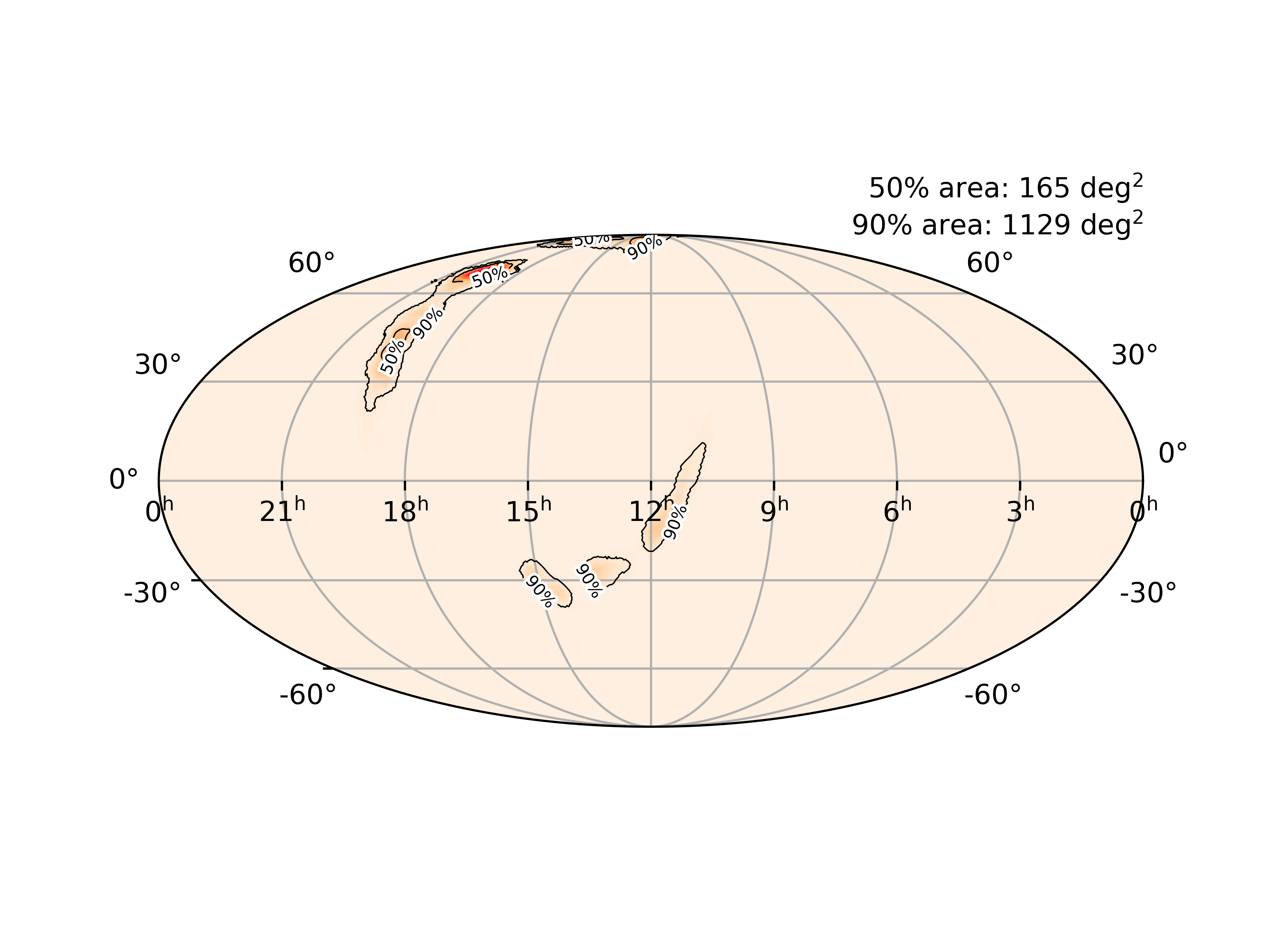}
\end{center} 
\caption{Sky localization probability maps (skymap) of three detectors for SSM200308. Mollweide all-sky projection in celestial (ICRS) coordinates. The outer and inner contours in the skymaps indicate the 90\% and 50\% credible regions respectively.}
\label{fig:skymap}
\end{figure}

The luminosity distance $d_L$ and inclination angle $\theta_{JN}$ posterior distributions are shown together in Fig.~\ref{fig:lalinference_Dist}. $\theta_{JN}$ corresponds to the angle between the system’s total angular momentum and the line of sight from the source to the observer. As expected from a CBC event, in the absence of observation of higher order modes, the two parameters are strongly correlated. The corner plot shows a clear bimodal distribution for $\theta_{JN}$ likely due to the fact that one cannot distinguish whether the system is being observed face-on ($\theta_{JN} \sim 0$) or face-away ($\theta_{JN} \sim \pi$), nevertheless the system being edge-on ($\theta_{JN} \sim \pi/2$) seems disfavoured. In Fig.~\ref{fig:skymap} we show the posterior sky localization of the event, which is relatively well localized, thanks to the three detectors being used in the PE analysis. \newline

The posteriors of the source of SSM200308 have converged to well-defined distributions that differ from their prior distribution (uniform in detector frame component masses, $\cos{\theta_{JN}}$ and power-law for $d_L$). Nevertheless, it is known that GW signals can be mimicked by Gaussian noise \cite{Morras:2022ysx} or non-Gaussian transients, especially given the relatively low SNR and high FAR.

    \begin{table}[t!]
	\begin{ruledtabular}
	\begin{tabular}{lr}
	   Parameter &   \\ \hline\\
	   Matched Filter SNR & \SNRmfPv  \\[4pt]
	   Primary mass ($M_\odot$)   & \PrimMassPv   \\[4pt] 
	   Secondary mass ($M_\odot$)  & \SecMassPv    \\[4pt] 
		Primary spin magnitude  & \PrimSpinPv    \\[4pt] 
		Secondary spin magnitude  &  \SecSpinPv      \\[4pt]
		Total mass ($M_\odot$)  & \MtotPv    \\[4pt]
        Detector-frame chirp mass ($M_\odot$) &  \ChirpPv \\ [4pt]
		Mass ratio $(m_2/m_1 \leq 1)$  & \qPv   \\[4pt]
		$\chi_{\text{eff}}$ \cite{PhysRevLett.106.241101,PhysRevD.82.064016} & \ChieffPv   \\[4pt] 
		$\chi_{\text{p}}$ \cite{PhysRevD.91.024043}  & \ChipPv   \\[4pt]
		Luminosity Distance (Mpc) & \DLPv  \\[4pt] 
		Redshift & \zPv  \\[4pt]
		$P( m_1< 1\ M_{\odot})$ & \ProbSSMonePv   \\[4pt]
        $P( m_2< 1\ M_{\odot})$ & \ProbSSMtwoPv   \\[4pt]
	\end{tabular}
	\end{ruledtabular}
	\caption{Parameters of the source of SSM200308. All masses are in the source frame. We assume $\textit{Planck15}$ Cosmology \cite{Planck15}. Errors quoted as $x^{y}_{z}$ represent the median of the marginalized one-dimensional posteriors, 5\% lower limit, and 95\% upper limit. 
	}
	\label{table:parameters}
\end{table}

\hspace{-10cm}
\section{Discussion}
\subsection{{Statistical tests}}
\label{sec:Discussion:statistical_test}

The LVK collaboration's search for SSM black hole binaries \cite{LIGOScientific:2022hai} {gives SSM200308 as one of the most} promising SSM candidates. With its false-alarm rate (FAR) of 0.20 yr$^{-1}$ and its global signal-to-noise ratio of $8.90$, the candidate is not far from, but still less significant than some confirmed events of O3b having similar SNR and FAR; e.g. GW200216\_220804 with FAR of 0.35 yr$^{-1}$ and SNR 9.40 or GW191230\_180458 with FAR 0.13 yr$^{-1}$ and SNR 10.30 \cite{LIGOScientific:2021djp}. {However}, one has to take into account that since SSM is a more speculative source of GWs, the significance required to claim a detection will be higher. \newline

One can define a Bayes factor that characterizes the model evidence, quoted $\mathcal{B} = Z_s/Z_n$, which is the evidence for the signal hypothesis divided by that for Gaussian noise. For the SSM200308 candidate, the natural logarithm of the Bayes factor given with our {coherent} parameter estimation {on the three detectors} is $\ln(\mathcal{B}_{H1L1V1}) = \texttt{\lnBHL}$.

In Ref.~\cite{Veitch:2009hd} a test {was introduced to} discriminate between a coherent signal {hypothesis} and incoherent {glitches that could mimic a} signal. The Bayesian coherence {test} ($\mathcal{B}_{\mathrm{coh},\mathrm{inc}}$) computes the odds between i) the hypothesis that a coherent CBC signal is present in the data and the hypothesis that instead, ii) the data presents gravitational-wave–like glitches occurring independently in each detector and mimicking a CBC signal: {$\ln{\mathcal{B}_{\mathrm{coh},\mathrm{inc}}} = \ln{\mathcal{B}_{H1L1V1}} - \ln{\mathcal{B}_{H1}} - \ln{\mathcal{B}_{L1}} - \ln{\mathcal{B}_{V1}}$}.
Using the coherent Bayes factor and the individual Bayes factors in each interferometer, listed in table~\ref{table:coh_inc_evidences}, we compute the $\mathcal{B}_{\mathrm{coh},\mathrm{inc}}$ for SSM200308 and find a value of $\ln{\mathcal{B}_{\mathrm{coh},\mathrm{inc}}}$ = \lnBcohinc, indicating a preference for the coherent {hypothesis} over the {incoherent} one. {Had the coherence test given a clear negative result, it could have been used to rule out the astrophysical signal hypothesis. However, its positive value does not imply that the signal is astrophysical. Methods have been developed to extend the coherence test to discern the astrophysical origin of a signal
(see Refs.~\cite{Isi:2018vst,2019PhRvD.100l3018A}), but neither of them are easily applicable to an SSM candidate due to the large computational cost and/or assumptions on the SSM population that have to be made. Extending these methods to SSM is left as future work.}

\begin{table}[t!]
    \bgroup
    \def\arraystretch{1.5}
	\begin{tabular}{| c | c | c | c | c |}
        \hline
    	$\ln{\mathcal{B}_{H1L1V1}}$ & $\ln{\mathcal{B}_{H1}}$ & $\ln{\mathcal{B}_{L1}}$ & 
        $\ln{\mathcal{B}_{V1}}$ & $\ln{\mathcal{B}_{\mathrm{coh},\mathrm{inc}}}$ \\ 
        \hline
        \lnBHL & \lnBH & \lnBL & \lnBV & \lnBcohinc \\
        \hline
    \end{tabular}
    \egroup
    \caption{\textbf{Natural logarithm} of the Bayes factors of the signal versus noise hypotheses obtained from the PE in the data of Hanford-Livingston $\ln{\mathcal{B}_{H1L1V1}}$, only Hanford/Livingston ($\ln{\mathcal{B}_{H1/L1}}$), only Virgo ($\ln{\mathcal{B}_{V1}}$), and the natural logarithm of the Bayes factor of the coherent versus incoherent hypothesis $\ln{\mathcal{B}_{\mathrm{coh, inc}}}$. For more discussion on the observed values of the Bayes factors, see Appendix~\ref{sec:BCR}
    }
\label{table:coh_inc_evidences}
\end{table}

\subsection{Nature of the source signal}
\label{sec:Discussion:nature_of_the_source_signal}

Given the results of the PE discussed so far, we can try to asses the possibilities for the nature of the source of SSM200308.

The neutron star nature of SSM200308 compact objects seems disfavored. Indeed, it is known from observations and simulations ~\cite{strobel_minimum_2001,bleeker_high-mass_2001,suwa_minimum_2018,nattila_neutron_2017} that it should be difficult to form neutron stars with masses below $\sim 0.9 M_\odot$ \footnote{However the possible observation of a neutron star mass as low as $0.77^{+0.20}_{-0.17} M_\odot$ was claimed in~\cite{2022NatAs.tmp..224D}.}. This is consistent with the detection of the first neutron star binary event GW170817 \cite{TheLIGOScientific:2017qsa} with source masses around $1.4 M_\odot$, and with the measured neutron star masses from binary pulsars~\cite{Ozel:2016oaf}.
Moreover, no known stellar evolution scenario can produce a black hole with a subsolar mass~\cite{Lattimer:2004sa}. 
Given the secondary component mass is firmly subsolar (having all the posterior distribution below 0.4$M_\odot$), if the signal does indeed come from a genuine astrophysical event, we are in the presence of compact objects from a new formation mechanism, alternative to the classic BH formation scenario.  

One can use the sensitivity volume values $\langle VT \rangle$ obtained in the O3 search of SSM objects~\cite{LIGOScientific:2022hai} to get an estimate on the rates for events similar to SSM200308, assuming that one event has been observed in the lowest chirp mass bin of Figure 1 of Ref.~\cite{LIGOScientific:2022hai}. We obtain merger rates between $400$ and $20000 \, {\rm Gpc^{-3} yr^{-1}}$ at 90\% C.L. for the \texttt{GstLAL} search in this bin, and similar numbers for other pipelines.  These rates are comparable or even higher than the inferred rates for vanilla neutron star mergers, but for objects that are expected to be outliers of the main neutron star population. Another origin should therefore be seriously considered.

The nature of the source signal remains an open question but there exist several theories about how such compact objects of subsolar mass might form. The subsolar origin of SSM200308's source could be explained by exotic compact objects such as boson stars \cite{schunck_topical_2003}, exotic black holes~\cite{Flores:2020drq,Singh:2020wiq,Kouvaris:2018wnh,PhysRevD.107.083012,PhysRevLett.126.141105} or by theoretical compact objects of primordial origin: Primordial Black Holes (PBHs) \cite{bird_did_2016, sasaki_primordial_2016, clesse_clustering_2017,Carr:2023tpt,LISACosmologyWorkingGroup:2023njw}. PBHs may have formed in the early Universe, shortly after inflation ended, from the direct collapse of highly overdense regions. Many studies show that the thermal history of the Universe can enhance the formation of PBHs during the Quantum Chromodynamic (QCD) phase transition (t $\sim 10\,\mu$s in the early universe)~\cite{Byrnes:2018clq,Carr:2019kxo} generating a distribution of PBH masses sharply peaked around one solar mass, leading to an enhanced rate of binary merger events in the sub-solar and solar mass range~\cite{Escriva:2022bwe,Franciolini:2022tfm}. The inferred characteristics of SSM200308, if truly coming from a GW event, would be consistent with the coalescence of two PBHs. 

\section{Conclusion}
In this work, we have performed, using ROQ methods, an in-depth analysis of one of the most significant candidates reported in the O3b search for SSM black hole binaries \cite{LIGOScientific:2022hai} with SNR = 8.90 and FAR = 0.20 yr$^{-1}$.
Even if the candidate does not show enough significance to claim the firm detection of a gravitational wave event, it is of great interest to study and characterize the candidate. We also demonstrate that the ROQ method can be efficiently used to reduce the computational cost of the PE for such long signals.
The inferred masses show that SSM200308, if coming from a GW event, is consistent with a binary of two SSM black holes; $m_1= \text{\PrimMassPv}\,M_{\odot}$ and $m_2 = \text{\SecMassPv}\,M_{\odot}$ (90\% credible intervals).
Given the very low masses of the candidate's components, their neutron star nature seems disfavoured~\cite{Ozel:2016oaf,Lattimer:2019mli}. The question of the nature of the source of SSM200308 therefore remains open.
The unusual characteristics of SSM200308 could be explained by the two components being black holes of primordial origin.
If SSM200308 is a real signal, we can expect that improved detector sensitivities and longer observing time would within a few years allow for the firm detection of an SSM black hole. 

\section{Acknowledgements}
We would like to thank Bhooshan Gadre and Viola Sordini for their work reviewing this paper within the LIGO and Virgo Collaborations respectively.
The authors acknowledge the use of the publicly available codes: \texttt{lalsuite} \cite{lalsuite} and \texttt{Bilby} \cite{Ashton:2018jfp}. 
They acknowledge support from the research project  PID2021-123012NB-C43 and the Spanish Research Agency (Agencia Estatal de Investigaci\'on) through the Grant IFT Centro de Excelencia Severo Ochoa No CEX2020-001007-S, funded by MCIN/AEI/10.13039/501100011033. 
GM acknowledges support from the Ministerio de Universidades through Grant No. FPU20/02857 
and JFNS acknowledges support from MCIN through Grant No. PRE2020-092571.  S.C. acknowledge support from the Belgian Francqui Foundation through a Francqui Start-up Grant, as well as the Belgian Fund for Research through a MIS and an IISN grants.  The authors are grateful for computational resources provided by the LIGO Laboratory and supported by the National Science Foundation Grants PHY-0757058 and PHY-0823459. 
This research has made use of data or software obtained from the Gravitational Wave Open Science Center \cite{Abbott:2019ebz, KAGRA:2023pio} (gw-openscience.org), a service of LIGO Laboratory, the LIGO Scientific Collaboration, the Virgo Collaboration, and KAGRA. LIGO Laboratory and Advanced LIGO are funded by the United States National Science Foundation (NSF) as well as the Science and Technology Facilities Council (STFC) of the United Kingdom, the Max-Planck-Society (MPS), and the State of Niedersachsen/Germany for support of the construction of Advanced LIGO and construction and operation of the GEO600 detector. Additional support for Advanced LIGO was provided by the Australian Research Council. Virgo is funded, through the European Gravitational Observatory (EGO), by the French Centre National de Recherche Scientifique (CNRS), the Italian Istituto Nazionale di Fisica Nucleare (INFN) and the Dutch Nikhef, with contributions by institutions from Belgium, Germany, Greece, Hungary, Ireland, Japan, Monaco, Poland, Portugal, Spain. The construction and operation of KAGRA are funded by Ministry of Education, Culture, Sports, Science and Technology (MEXT), and Japan Society for the Promotion of Science (JSPS), National Research Foundation (NRF) and Ministry of Science and ICT (MSIT) in Korea, Academia Sinica (AS) and the Ministry of Science and Technology (MoST) in Taiwan.

\appendix

\section{Priors}
\label{sec:priors}
The PE results were performed using default priors that are intended not to make strong assumptions about the source properties. We choose a prior that is uniform in redshifted component masses and in spin magnitudes, isotropic in spin orientations and sky location. The priors that were used are detailed in Table \ref{table:priors}.

\begin{table*}[htbp]
    \centering
    \caption{Parameter Priors}
    \begin{tabular}{|c|c|c|}
        \hline
        \textbf{Parameter} & \textbf{Prior Type} & \textbf{Range} \\
        \hline
        Chirp Mass ($\mathcal{M}$) & UniformInComponentsChirpMass & $[0.351, 0.355] \, M_{\odot}$ \\
        Mass Ratio ($q$) & UniformInComponentsMassRatio & $[0.1, 1.0]$  \\
        Mass 1 ($m_1$) & Constraint & $[0.142, 10] \, M_{\odot}$  \\
        Mass 2 ($m_1$) & Constraint & $[0.142, 10]\, M_{\odot}$ \\
        Declination (DEC) & Cosine & $[-\pi/2, \pi/2]$ \\
        Right Ascension (RA) & Uniform (Periodic) & $[0, 2\pi]$ \\
        $\cos\theta_{JN}$ & Uniform & $[-1, 1]$ \\
        Polarization Angle ($\psi$) & Uniform (Periodic) & $[0, \pi]$ \\
        Phase ($\phi$) & Uniform (Periodic) & $[0, 2\pi]$ \\
        Spin Parameter $a_1, a_2$ & Uniform & $[0, 0.8]$ \\
        Tilt ($\theta_1$,$\theta_2$) & Sine &  $[0, \pi]$ \\
        Difference between azimuthal spin angles $(\phi_{12})$ & Uniform (Periodic) & $[0, 2\pi]$ \\
        Phase between orbital and total angular momenta: $(\phi_{jl})$ & Uniform (Periodic) & $[0, 2\pi]$ \\
        Luminosity Distance ($d_L$) & PowerLaw & $[5, 300] \, \text{Mpc}$ \\
        \hline
    \end{tabular}
    \label{table:priors}
\end{table*}

\section{Discussion on the $\mathcal{B}_{\mathrm{coh},\mathrm{inc}}$}
\label{sec:BCR}

\begin{figure*}[t!]
\begin{center}
\includegraphics[width=\textwidth]{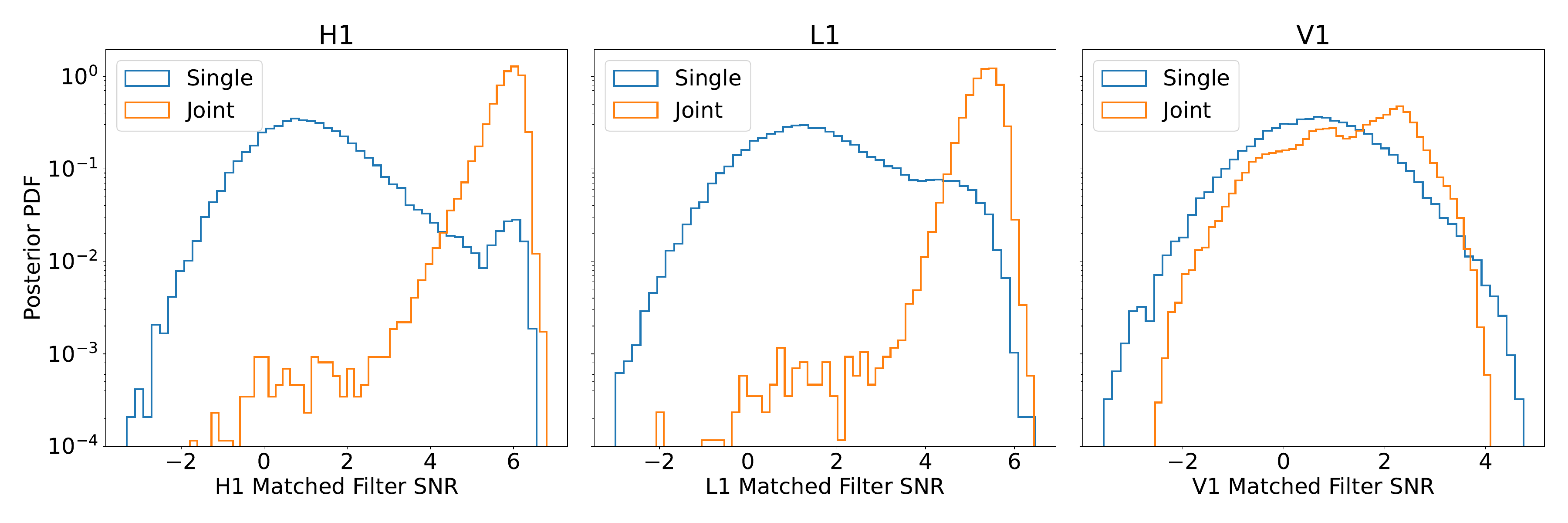}
\end{center} 
\caption{Posterior probability density function (PDF) of the matched filter SNR in each detector, both for the PE using the data of only that (Single) detector and of all (Joint) detectors at the same time.}
\label{fig:single_vs_joint_mf_SNR}
\end{figure*}

Looking at table~\ref{table:coh_inc_evidences}, we note that the single detector Bayes factors are very close to zero, and therefore, the $\mathcal{B}_{\mathrm{coh},\mathrm{inc}}$  is very close to the Bayes factor of the joint analysis ($\ln{\mathcal{B}_{\mathrm{coh},\mathrm{inc}}} \sim \ln{\mathcal{B}_{H1L1V1}}$). At first glance, it might seem this is because the PE does not find the signal in the single detector analyses, and therefore we are getting a skewed value of $\ln{\mathcal{B}_{\mathrm{coh},\mathrm{inc}}}$. However, looking at Fig.~\ref{fig:single_vs_joint_mf_SNR}, we can observe that the single detector analyses finds maximum values of the SNRs very close the maximum values found by the joint analysis.

The reason why the single detector analyses have such small Bayes factors is that the Likelihood is related to $\exp(\mathrm{SNR}^2/2)$ and the SNR is not large enough to make the posterior significantly different from the prior and thus a value of $\ln(\mathcal{B})$ significantly larger than 0. However, in the joint analysis, the matched filter SNR has a value of $\sim 8$ and the Likelihood becomes large enough to ``fight'' against the prior and give the large value of the Bayes factor listed in table~\ref{table:coh_inc_evidences}.

The network matched filter SNR, and consequently the coherent Bayes Factor, are so large because the tails of the SNR distributions of Fig.~\ref{fig:single_vs_joint_mf_SNR} correspond to the same model parameters, with the correct phases and timings to be added constructively. Additionally, removing the condition that there has to be coherence between detectors, and giving more freedom to the signal in the single detector analyses, does not lead to significantly larger single detector SNRs than those seen by the joint analysis.

\bibliographystyle{apsrev4-1}
\bibliography{Biblio}

\end{document}